\documentclass[12pt]{spieman}  %>>> 12pt font mandatory; use this for US letter paper 
\usepackage[]{graphicx}
\usepackage{setspace}
\usepackage{tocloft}

\newcommand{\arcsec}{\ensuremath{^{\prime\prime}}}

\title{WFIRST-AFTA Coronagraphic Operations: Lessons Learned from the Hubble Space Telescope and the James Webb Space Telescope} 

\author{John H. Debes\supscr{a}, Marie Ygouf\supscr{a}, Elodie Choquet\supscr{a}, Dean C. Hines\supscr{a}, Marshall Perrin\supscr{a}, David A. Golimowski\supscr{a}, Charles-Phillipe Lajoie\supscr{a}, Johan Mazoyer\supscr{a}, Laurent Pueyo \supscr{a}, R\'{e}mi Soummer \supscr{a}, Roeland van der Marel\supscr{a}}

\affiliation{\supscrsm{a}Space Telescope Science Institute, 3700 San Martin Dr., Baltimore, MD, USA, 21212}
\cftpagenumbersoff{figure}
\cftpagenumbersoff{table} 
\begin{document} 
\maketitle 

%%%%%%%%%%%%%%%%%%%%%%%%%%%%%%%%%%%%%%%%%%%%%%%%%%%%%%%%%%%%% 
\begin{abstract}
The coronagraphic instrument currently proposed for the WFIRST-AFTA mission will be the first example of a space-based coronagraph optimized for extremely high contrasts that are required for the direct imaging of exoplanets reflecting the light of their host star.  While the design of this instrument is still in progress, this early stage of development is a particularly beneficial time to consider the operation of such an instrument.  In this paper, we review current or planned operations on the Hubble Space Telescope (HST) and the James Webb Space Telescope (JWST) with a focus on which operational aspects will have relevance to the planned WFIRST-AFTA coronagraphic instrument.  We identify five key aspects of operations that will require attention: 1) detector health and evolution, 2) wavefront control, 3) observing strategies/post-processing, 4) astrometric precision/target acquisition, and 5) polarimetry.  We make suggestions on a path forward for each of these items.
\end{abstract}

%>>>> Include a list of up to six keywords after the abstract

%>>>> Include contact information for corresponding author
{\noindent \footnotesize{\bf Address all correspondence to}: John Debes, Space Telescope Science Institute, 3700 San Martin Dr., Baltimore, MD 21212, USA; Tel: +1 410-338-4782; Fax: +1 410-338-5090; E-mail:  \linkable{debes@stsci.edu} }
%%%%%%%%%%%%%%%%%%%%%%%%%%%%%%%%%%%%%%%%%%%%%%%%%%%%%%%%%%%%%

\begin{spacing}{2}   % use double spacing for rest of manuscript

%%%%%%%%%%%%%%%%%%%%%%%%%%%%%%%%%%%%%%%%%%%%%%%%%%%%%%%%%%%%%
\section{Introduction}
\label{sect:intro}  % \label{} allows reference to this section
The WFIRST-AFTA coronagraphic instrument (CGI) enables ground-breaking exoplanet discoveries using optimized space-based coronagraphic high contrast imaging.  Its goal of 10$^{-9}$ contrast at an inner working angle of %0\farcs1 
0.1\arcsec\,relative to an astrophysical point source will result in the direct detection of exoplanet candidates around nearby stars that are directly reflecting light from their hosts and push to within an order of magnitude of the requirements for a large aperture space telescope to directly image nearby terrestrial planets.  The CGI design currently consists of a Hybrid Lyot Coronagraph (HLC) mode that will be primarily used for planet detection.  There will also be a Shaped Pupil Coronagraph (SPC) mode, which will have more moderate contrast and a smaller field of view.  The SPC coupled with an integral field spectrograph (IFS) mode will provide spectral characterization of exoplanets in addition to direct imaging.  Both imaging and spectroscopic modes will use Electron Multiplying Charge Coupled Devices (EMCCDs)\cite{wilkins14}, which provide high sensitivity for faint sources in the optical bandpass.

Achieving the goals of the CGI requires not only a sound design for launch, but a comprehensive and complete plan for operating the instrument and adapting to its changes throughout the duration of the mission. The present manuscript is aimed at highlighting the synergies between HST, JWST, and WFIRST-AFTA in coronagraphic operations. It relies on information regarding the instrument presented in the 2015 SDT report \cite{spergel15}.  Aspects discussed here are lessons learned from 15+ years of coronagraphy with HST and plans for JWST.  HST has had optical and Near-IR coronagraphic capabilities in space for most of its lifetime, while JWST will represent the first NASA flagship observatory with a diverse suite of various high contrast imaging approaches.  The operation of the coronagraphs for each mission have direct ties to the WFIRST-AFTA CGI.

%%%%%%%%%%%%%%%%%%%%%%%%%%%%%%%%%%%%%%%%%%%%%%%%%%%%%%%%%%%%%
\section{Hubble Space Telescope Operations} 
\label{sec:HST}
Over its 25-year lifetime, HST has had four separate instruments with high contrast coronagraphic imaging capability, though none of which were specifically optimized for high contrast imaging at extreme inner working angles.  As such, HST's high contrast instruments have achieved high contrast (such as the 10$^{-9}$ contrast detection of Fomalhaut b at r=9\arcsec) or small inner working angles (0.25\arcsec\ for STIS), but not both simultaneously.  This is primarily because there is no instrumentation that is capable of significantly suppressing diffracted starlight from the wings of the telescope's point spread function, which are more severe from mid-frequency wavefront errors on the primary mirror.  One of the four original instruments on Hubble, the Faint Object Camera, had occulting fingers that were not highly used \cite{nota}.  With the second servicing mission of HST the Space Telescope Imaging Spectrograph (STIS)\cite{woodgate98} and Near-infrared Imager and Multi-Object Spectrometer (NICMOS)\cite{thompson92} were installed, which in concert with Reference Difference Imaging (RDI) and Angular Differential Imaging (ADI) allowed for high quality, high contrast images of faint companions, circumstellar debris disks, and protoplanetary disks at the level of 10$^{-4}$ contrast at inner working angles of 0.3\arcsec\ \cite{lowrance99,schneider99,grady99}.  High contrast images of other astrophysical objects, such as the host galaxies of quasars were also imaged \cite{hines99}.  While NICMOS is no longer operational, archival coronagraphic data is still being utilized in concert with modern post-processing techniques, which we discuss further in Section \ref{sec:post}.  STIS possesses two focal plane wedges and two occulting bars with a slightly undersized Lyot stop in the pupil plane.  NICMOS possessed a hole bored into a mirror that directed light to its second camera which served as an occulting spot as well as a cold mask that acted as a crude Lyot stop.  The High Resolution Channel (HRC) on the Advanced Camera for Surveys (ACS)\cite{ford98}, launched during servicing mission 3B, possessed two coronagraphic spots with inner working angles of %0\farcs9 
0.9\arcsec\,and %1\farcs8
1.8\arcsec, which provided high contrast, deep observations of many disks across broadband optical filters \cite{kalas05,golimowski06}.

At present, STIS is the only remaining operational high contrast instrument on board HST and in the case of inner working angle, detector size, and passband the most similar to the WFIRST-AFTA CGI design.  Its coronagraphic mode is chromatically unfiltered and follows the quantum efficiency of its CCD detector, which is sensitive to a broad UV-through Near-IR bandpass spanning 200-1000~nm.  Its newest commissioned position for high contrast imaging is located on the bent occulting finger\footnote{http://www.stsci.edu/hst/stis/strategies/pushing/coronagraphic\_bars}, which has a width of 300~mas, and thus an inner working angle for solar-type stars that corresponds to 3$\lambda$/D at an effective wavelength of 630~nm.  This small inner working angle location began routine operations for Cycle 22 in 2014.  

\begin{figure}
   \begin{center}
   \begin{tabular}{c}
   \includegraphics[height=5.5cm]{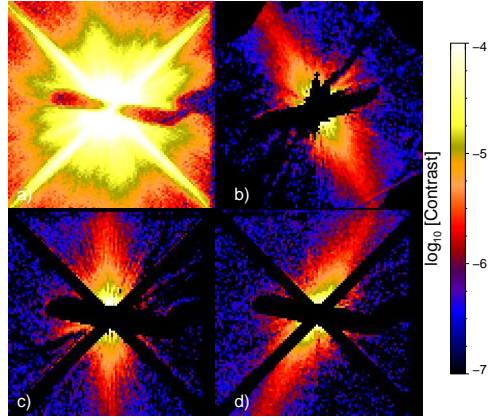}
   \end{tabular}
   \end{center}
   \caption 
   { \label{fig:f1} %>> use \label inside caption to get Fig. number with \ref{}
STIS coronagraphic performance using the bent finger occulter.  a) Raw log-scale image of $\beta$ Pictoris  (image size is %2\farcs5$\times$2\farcs5
2.5\arcsec$\times$2.5\arcsec).  b) Final, log-scale image of the debris disk oriented so that North is up and East to the left.  The overlapping masked regions combine to create an effective inner working angle of 0.25\arcsec, or 4.75~AU at the distance of $\beta$ Pictoris.  The image was created from the subtraction of a near-contemporaneous reference PSF, and the finger and diffraction spikes were masked
c) single spacecraft orientation image of the debris disk in orbit around $\beta$ Pictoris.  d) second single spacecraft orientation image of the $\beta$-Pictoris debris disk.  The two orientations are combined and mask regions are treated as missing data to create a final image of the disk.  } 
   \end{figure}

Initial results from commissioning data of GO~12923 (PI:Gaspar) are promising.  The occulted PSF is not suppressed by the finger, but by observing a contemporaneous reference PSF and observing at two or more spacecraft orientations.  Achieved contrasts relative to the peak of the host star PSF of better than $\sim10^{-4}$ per pixel at 250~mas have been obtained (a factor of 10 improvement over the raw contrast--See Figure \ref{fig:f1}).  With an optimal execution of 3-6 spacecraft orientations, this might be improved to the half-width of the finger (150~mas).  This visible high contrast capability is complementary to ground-based extreme AO systems that primarily work in the Near-IR, and is particularly attractive for sources too faint for groundbased AO or for highly axisymmetric structures that are not well-recovered with post-processing.  Several science programs using this bent finger have now been executed.

We investigate useful overlaps between HST and CGI operationally on the front end (target acquisition, jitter, etc.), calibration (flux calibration, flat-fielding), and back-end stages of data taking (pipeline produced products, post-processing).  In particular, we focus on links to the health and monitoring of the proposed detectors for the CGI, and mask repeatability.  Finally, post-processing techniques that have been successfully applied to many ground-based AO systems are filtering back to space-based coronagraphy \cite{lafreniere07,soummer12}; we explore what has been done on HST and how that may be applied to higher contrasts required for the WFIRST-AFTA CGI.

\subsection{Detector Health and Monitoring}
\label{sec:monitoring}
\begin{figure}
   \begin{center}
   \begin{tabular}{c}
   \includegraphics[width=10cm]{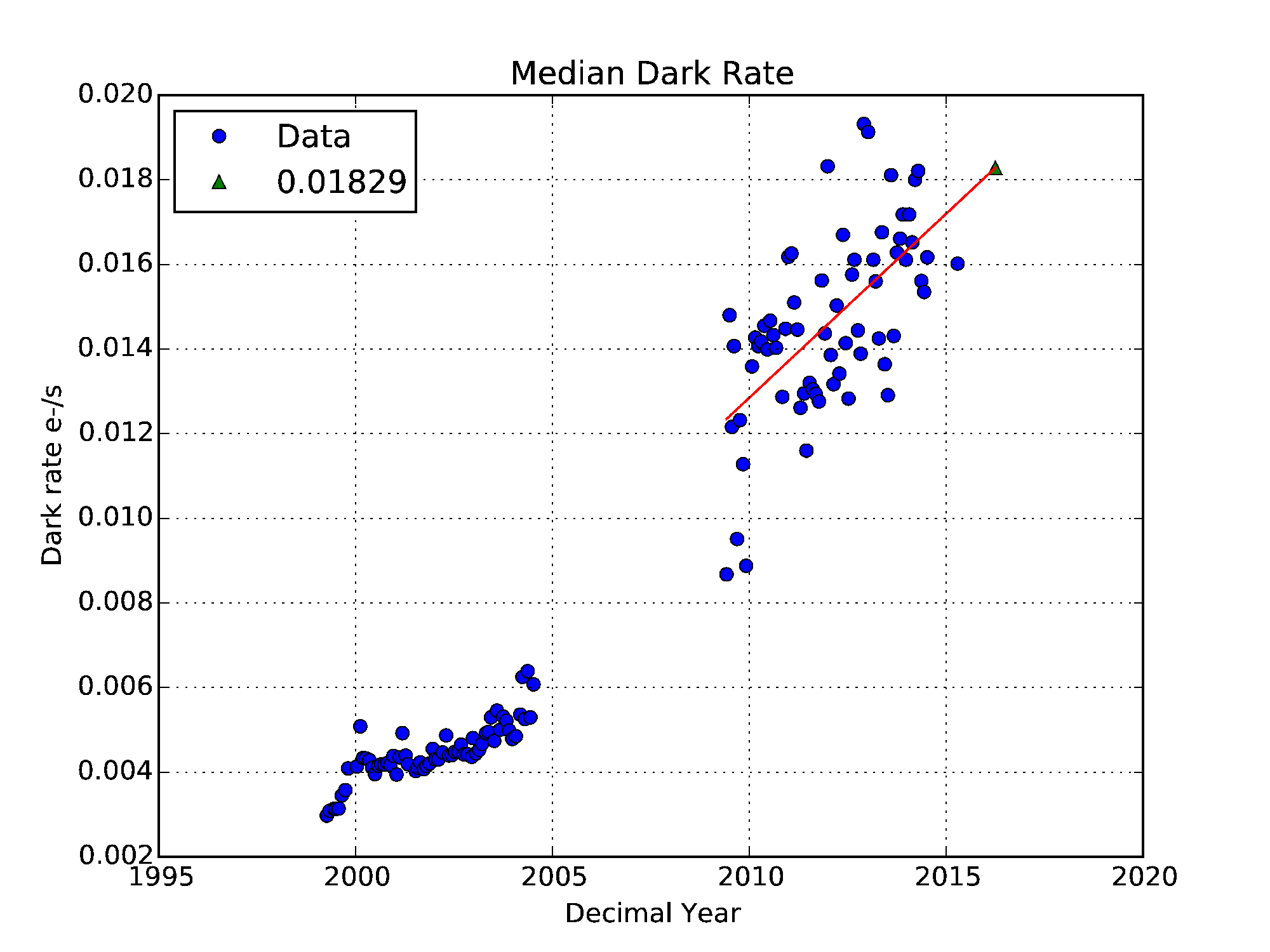}
   \end{tabular}
   \end{center}
   \caption 
   { \label{fig:f2} %>> use \label inside caption to get Fig. number with \ref{}
Plot of STIS dark rate vs. time. The CCD dark rate and CTE are monitored monthly and semi-annually on STIS respectively.} 
   \end{figure}

The basic detector technology of CCDs is sufficiently similar to the current proposed EMCCD technology \cite{wilkins14} that it is worth investigating the evolution of an EMCCD during a 5-year mission.  The expected count rate of an exoplanet's photons onto the WFIRST-AFTA CGI is on the order of $\ll$1 s$^{-1}$, not only requiring absolute stability of the high contrast dark hole, but extremely low dark noise and read noise.  It will therefore not be sufficient to have a detector that launches with the required noise levels, but one that continues to meet its requirements over the nominal operational baseline.  The harsh radiation environment of space imposes severe conditions on a CCD detector, primarily from impacts of charged particles with the detector.  The result of these impacts is an increase of dark rate and hot pixels, which has also been measured with the radiation testing of EMCCDs \cite{smith06}.  In that test, EMCCD dark rates increased by roughly a factor of two after a total fluence of 2$\times10^{10}$ protons cm$^{-2}$ was applied for 95 separate detectors.  Hot pixel number increased as well, similar to what is observed in regular space-based CCDs.  Current HST operations include monitoring of read noise, dark current, and charge transfer inefficiency (See Figure \ref{fig:f2} for the STIS dark rate vs. time).

Concurrent with these effects is the degradation of charge transfer efficiency (CTE) in space-based CCDs on HST \cite{anderson10}.  Radiation damage creates ``charge traps'' which release electrons on timescales longer than a pixel-to-pixel charge transfer, resulting in residual charge being readout in rows and columns further from the readout amplifier.  Fractionally more traps affect smaller charge packets, such that sources with low numbers of counts on a detector can be lost completely.  Pixels with high numbers of counts (either from hot pixels, the wings of a PSF, or the core of a bright star) create charge trails that point away from the amplifier direction on the detector.  The brightest pixel will regulate the exact profile of a charge trail.  All of these effects can decrease the signal-to-noise of faint objects, change extended source morphology, degrade astrometric precision, or present potential problems with post-processing, as the PSF structure will be dependent on the number of counts on the detector.  EMCCDs are also sensitive to CTE degradation, with a 25 e$^-$ signal at the center of the detector with a 35 e$^-$ pre-flash signal applied being completely lost after a total fluence of 2$\times10^{10}$ protons cm$^{-2}$ was applied \cite{smith06}.  This proton fluence is roughly equivalent to ten times the expected dose over a six-year mission duration in an L2 orbit, and presents a worst case scenario.  Even so, the degradation in SNR for faint signals happens with smaller radiation levels, which has been observed for all of the HST instruments.

Using the performance of HST CCDs, we can estimate the level to which trailing may be an issue for coronagraphy using the empirical estimates of CTE degradation for ACS and STIS under the assumption that the coronagraphic dark hole is located in the middle of the 1k$\times$1k EMCCD detector.  These estimates are no better than an order of magnitude, but should provide a framework for future testing of EMCCDs.  

We assume that the tail structure due to imperfect CTE  means that a majority of this charge is released within a handful of pixels--the trails from point sources with  $\sim$10$^4$ counts on the detector are $\sim$0.1\% of the peak 20 pixels away\cite{anderson10}.  As an example, consider Figure 3-26 from Ref. \citenum{spergel15} (reproduced in Figure \ref{fig:f3b}), which shows the contrast achieved for the HLC.  Assuming a single readout will avoid saturating the detector within the core, one can assume that the central region will have on order 10$^4$ counts.  The charge trapped tail from this central portion would be higher than 10 counts at %0\farcs2
0.2\arcsec, well outside the IWA of the coronagraph and degrading the SNR for any putative planet in that portion of the dark hole.   The high contrast dark hole generally features a bright outer edge that might impact this effect as well: while in exo-planet mode the CGI will most certainly feature a  field-stop-like focal plane that will prevent bright starlight from this edge to reach the detector. However this field stop might be detrimental to disk science and it is not clear yet if all observations will be carried out using it.  Similarly for the IFS, masking of the outer portions of the shaped pupil coronagraph (SPC) will be a compromise between discovery space and leakage of the exterior PSF.  Wherever the PSF leaks through the hard stop masking, trailing will occur at varying levels, some of which may impact final SNR.  

There are several ways to mitigate the effects of CTE degradation on the HST optical CCDs, including pre-flashing \cite{smith06}, charge injection, and pixel-based CTE corrections \cite{anderson10}.  None of these options is desirable for the CGI, since they fail to mitigate the degradation in the source signal, and more often than not increase the noise associated with an observation either through increase counts due to a background (pre-flashing, charge injection), or through the enhancement of noise (pixel-based CTE correction).  A simpler approach for the highest contrast applications would be to favor the placement of the dark hole nearest to the amplifier(s) of the EMCCD.  This is similar to the ``E1'' spectroscopic location for most slits in STIS, where spectroscopic point sources are placed 100 pixels away from the STIS amplifier to decrease the number of needless readout transfers.  Since the effects of CTE degradation are proportional to $(CTE)^y$, where $y$ is the number of rows away from the amplifier, minimizing y decreases the amount of deferred charge.  For example, after 5 years of operation, the CTE per transfer of STIS for a point source with 150 counts was on the order of 0.99965 per transfer, meaning that in the middle of the detector it lost 17\% of its flux to CTE degradation\cite{goudfrooij06}.  Conversely, the same point source 100 pixels from the readout amplifier would only lose 6\%, a factor of 3 lower.  

\begin{figure}
   \begin{center}
   \begin{tabular}{c}
   \includegraphics[width=10cm]{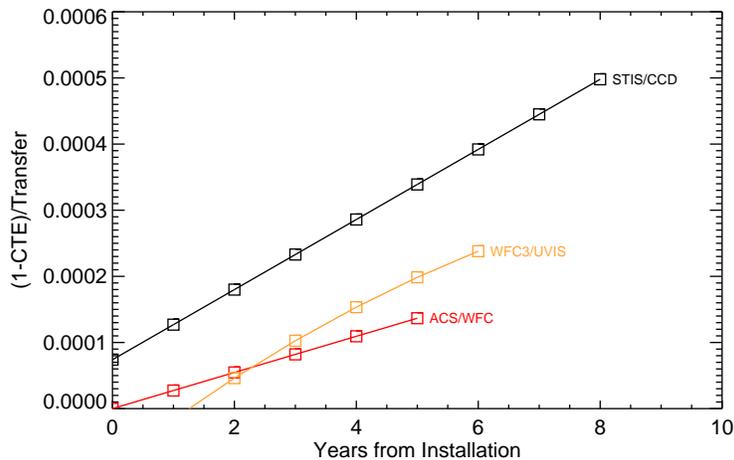}
   \end{tabular}
   \end{center}
   \caption 
   { \label{fig:f3} %>> use \label inside caption to get Fig. number with \ref{}
Comparison of charge transfer efficiency (CTE) per pixel transfer as a function of time of operation for STIS, ACS, and WFC3.  The relations are derived from empirical fits to CTE monitoring \cite{goudfrooij06,chiaberge09,baggett15}.  Differences in CTE evolution are partly anti-correlated to solar activity\cite{baggett15}.  During solar minima, charged particle impacts from the South Atlantic Anomaly are more frequent.  For STIS and ACS these relations held for early operations before their shutdowns in 2004 and 2007 respectively.  Post-SM4 CTE evolution of ACS in particular has changed since Servicing Mission 4.  Since WFIRST-AFTA will be in a higher orbit, it may see a more steady flux of radiation than HST.} 
   \end{figure}

\begin{figure}
   \begin{center}
   \begin{tabular}{c}
   \includegraphics[width=10cm]{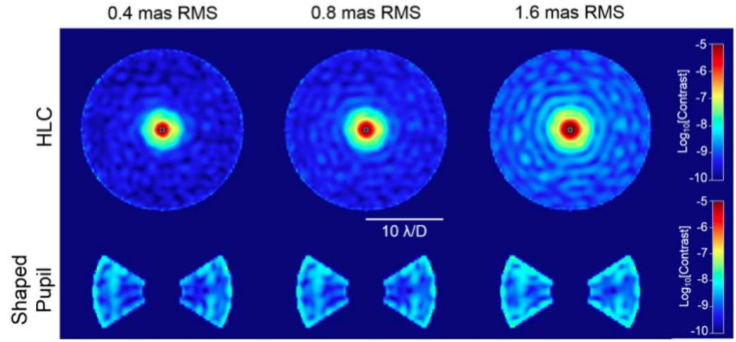}
   \end{tabular}
   \end{center}
   \caption 
   { \label{fig:f3b} %>> use \label inside caption to get Fig. number with \ref{}
 Figure showing the unmasked residual light levels from the HLC, reprinted from Figure 3-26 of the WFIRST-AFTA 2015 SDT Report.  Bright regions on the detector, such as from the brighter central spot in the HLC, can create trailing due to CTE degradation.} 
   \end{figure}

Operation of the HST CCDs in space suggests that CTE degradation is mostly independent of initial conditions of a detector and more a function of the solar radiation flux during a mission.  Pre-launch radiation tests of ACS CCDs showed that initial CTE was not a dominant determinator of CTE degradation due to radiation.  Both STIS and ACS SITe CCDs were fabricated many years ago and their CTE has degraded roughly linearly in time  (See Figure \ref{fig:f3}).  The more modern WFC3 UVIS instrument is an e2v Ltd. thinned, back-illuminated CCD and  also suffers from CTE degradation at roughly similar rates to ACS \cite{baggett15}.  After five years of operation the UVIS CCD has a similar magnitude of CTE degradation as to STIS in 2002 to within a factor of 2 and ACS in 2007 to within 30\%.  WFC3's rate of CTE degradation was initially more rapid than that of ACS in orbit, which was primarily due to a locally higher radiation dose at WFC3 installation than was present for ACS' initial years of operation.  Predictions for WFC3's CTE degradation exist--the slope of predicted CTE degradation of the WFC3 detectors was found to be -3.4$\times10^{-5}$~yr$^{-1}$\cite{waz01}, while the actual slope as measured on-orbit is closer to -5$\times10^{-5}$~yr$^{-1}$.  Both WFC3 and STIS have had similar CTE degradation slopes over the initial 6-years of operation, while ACS suffered from a smaller rate of degradation (2.7$\times10^{-5}$~yr$^{-1}$) over its first six years, which seems correlated with the Solar Cycle\cite{baggett15}.       

CTE degradation can have a myriad of unforseen effects on the final SNR of a high contrast imaging strategies--we therefore strongly advocate for significant testing of the detailed behavior of EMCCDs in the face of CTE degradation in the laboratory, with particular attention to what sources of uncertainty in dosage (such as due to the solar cycle) may have on CTE degradation rates.  CCD instruments onboard HST have utilized long dark images and the behavior of trailing from hot pixels to create empirically derived models of both charge degradation and trail brightness as a function of distance from a source \cite{anderson10}.  CTE degradation in STIS is monitored via observations that switch amplifiers on the detector to change the effective $y$ position of a lamp source to quantify the amount of charge lost on the detector due to CTE degradation \cite{goudfrooij06}.  These types of tests can be implemented on irradiated EMCCDs and tested on low- and high-gain modes to determine a quantitative impact for specific CGI designs as a function of total irradiation dose.  Models based on these results can then be applied to simulated data to investigate the impact to high contrast imaging after the start of operations.

\subsection{Repeatability of Mask Deployment}
\label{sec:repeat}
The alignment of masks in the CGI will be critical to its final contrast performance.  While the CGI will have the Low Order WaveFront Sensing and Control (LOWFSC)~\cite{Wallace-p-2011} and fine steering mirror (FSM), secular drifts of masks will need to be monitored and accounted for in target acquisition or wavefront sensing, especially for the HLC.  Similarly, the SPC occulting masks will need to be properly aligned on the detector relative to the PSF.  

All of HST's high contrast imaging instruments suffered from secular drifts or non-repeatabilities of pupil mask or occulter components.  The secular drifts of the coronagraphic spots in the ACS HRC were random and varied across 4 pixels on weekly timescales($\sim$160~mas) \cite{krist04}.  Non-repeatabilities of the coronagraphic mask deployment were on the order of 12~mas\cite{krist04}.  In most cases, these drifts did not impact performance, but did require monitoring to ensure accurate centering of a target behind the mask. 

STIS also suffers from non-repeatability in its acquisition pointing, slit wheel, and mode select mechanism, which controls the relative placement of the aperture (50CORON) that contains the occulting wedges and bars relative to the CCD detector \cite{kinney95,downes97,downes97msm}.  The magnitude of this non-repeatability is roughly 13~mas, primarily in the y-direction on the detector and along the line of movement for the slit wheel.  The non-repeatability causes two impediments to high contrast: 1) at one edge of an occulting bar, brighter regions of the PSF ``peek'' out and can saturate the detector, thus limiting the effective IWA of an observation; 2) non-repeatability also limits the utility of reference PSF stars at the IWA for classical PSF subtraction.  No measured secular drifts have been reported for the coronagraphic aperture masks.  Repeatability for STIS was coarsely measured on the ground and the requirement for the repeatability of the instrument components was on the order of  20~mas.

NICMOS suffered secular drifts in its coronagraphic hole 0.25 pixels (20~mas) on the timescale of 3 orbits, and 1 pixel on the timescale of days (75~mas), with a total movement of +2 pixels (150~mas) in the horizontal, and +5 pixels (375~mas) in the vertical on the detector over 7 years of operation \cite{schultz04,schultz03}.  Additionally, the cold mask (which acted as a crude Lyot Pupil Mask for the NICMOS coronagraph) had shifted throughout the lifetime of NICMOS, which introduced non-repeatabilities in PSF structure that degraded PSF subtraction performance\cite{krist98}.  The magnitude of this drift was roughly 1.4\% of the pupil radius over 2 years, and was not expected based on ground testing \cite{krist99}.  The cold mask and hole shifts were primarily due to the NICMOS dewar anomaly \cite{dewar99} that significantly changed the optical alignment of various instrument components.  The hole shifts continued to show trends after the NICMOS cryocooler was installed, presumably because of additional, slow movements of the NICMOS optical bench or optics \cite{schultz9906}.  Shifts in the cold masks directly limited the contrast achieveable for classical PSF subtraction when observations were separated by long times.  

The slit and mask wheels of the CGI will need to be tested, and their repeatability should be at a level similar to the LOS jitter in order to minimize the risk to contrast performance--these requirements are more stringent than those put in place for HST.  Furthermore, the repeatability and positioning will need to be monitored throughout the mission lifetime to ensure that unforseen behavior is not missed.

%\section{Operations and post-processing: lessons from HST and preliminary work using  WFIRST-AFTA simulated data}\label{sec:process}
\section{Post-processing: lessons from HST and preliminary work using  WFIRST-AFTA simulated data}\label{sec:process}
\label{sec:post}

Calibration of the post-coronagraph PSF will play a major role in the WFIRST-AFTA CGI instrument  \cite{spergel15}. In the context of optimizing the scientific return of HST, significant work has been carried out in the past few years in order to analyze coronagraph data obtained using the RDI and ADI observing strategies. 

\subsection{PSF subtraction techniques}\label{reduction}
\label{sec:psfsub}
The highest contrast levels in space with STIS have been achieved by simultaneously including a reference PSF star (i.e. reference differential imaging; RDI) in addition to observing a target system at multiple spacecraft orientations (i.e., azimuthal differential imaging; ADI)~\cite{Schneider-a-2014}.  Multiple spacecraft orientations also allows for post-processing the data by iteratively subtracting off static residuals seen across observations\cite{Stark-a-2014}.  Recently, post-processing algorithms that rely on a large library of reference PSFs, such as the Locally Optimized Combination of Images (LOCI)~\cite{Lafreniere-a-07b}, and of their principal components~\cite{Soummer-a-2012,Amara-a-2012}, have emerged in the context of ground based coronagraphy. 

\begin{figure}[!h]  
    \centering
       \includegraphics[width = 10cm]{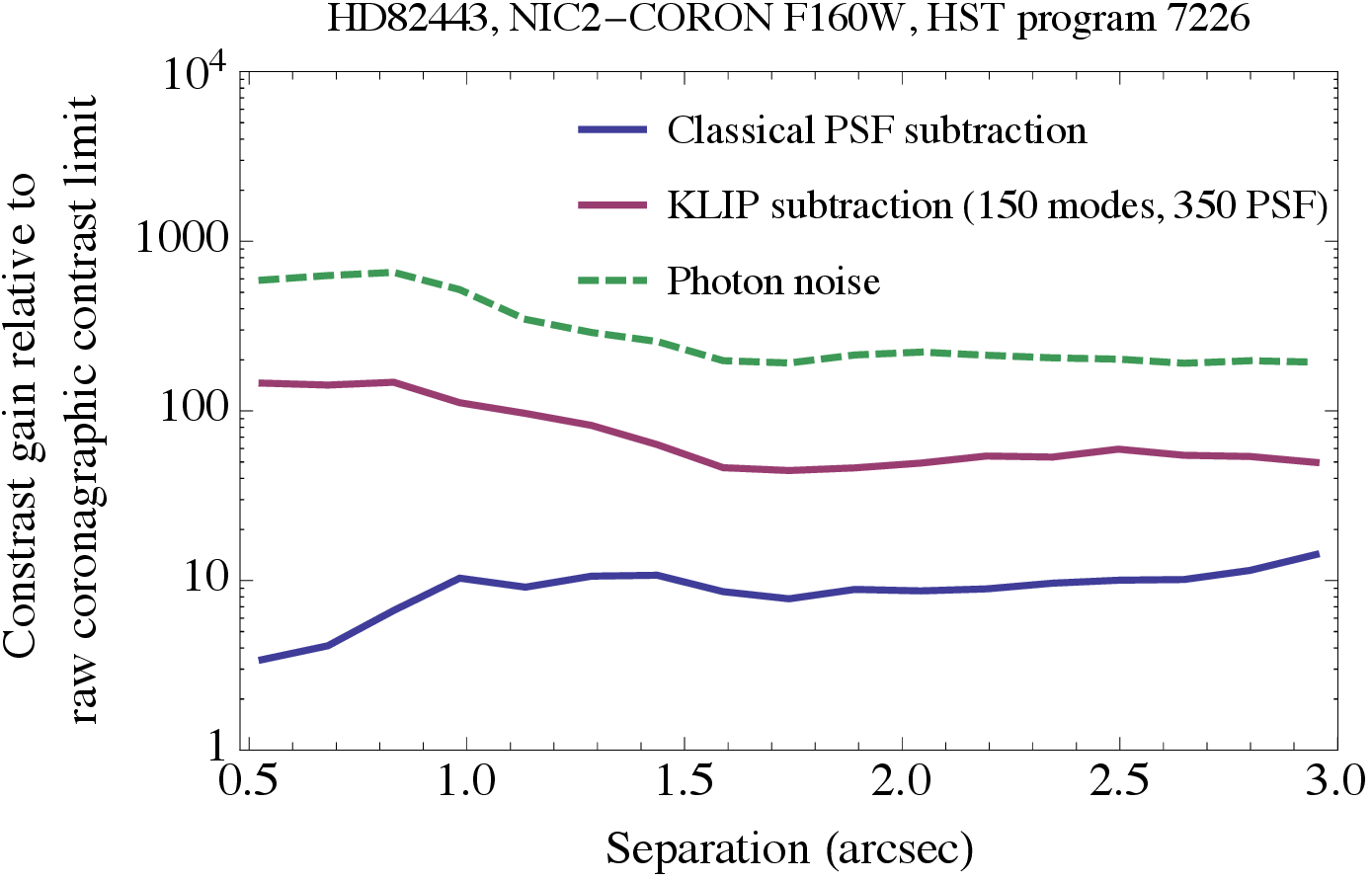}
    \caption{\label{ALICEFigure} Example of contrast gain curves relative to the raw coronagraphic contrast limit as a function of separation obtained with a classical PSF subtraction (blue) and a KLIP (red) reduction. The red curve does not take into account the KLIP throughput.}
\end{figure}

These techniques have been applied to the large NICMOS coronagraph observational history and yielded unprecedented images of exo-planets and debris disks~\cite{Soummer-a-2011,Soummer-a-2014,Rajan-p-2015a}. The Archival Legacy Imaging of Circumstellar Environments (ALICE; HST-AR-12652, PI: Soummer) project relies on these algorithms and is currently mining the entire NICMOS archive: Figure~\ref{ALICEFigure} illustrates the typical contrast gain obtained using very large (350) NICMOS PSF libraries, with a final contrast close to the photon noise limit. In parallel this technique has also been applied to HST-WFC3~\cite{Rajan-p-2015a}.  Similar post-processing is being actively considered for JWST as well (See Section \ref{sec:dput} for more details).

Inclusion of reference PSFs allow sampling of the PSF wings over a variety of telescope or pointing states.  While it is desired to have the exact thermal, pointing, and wavefront state on the telescope, a main assumption with enhancing contrast on HST and JWST is that a coronagraph's PSF response to small variations in wavefront, pointing, and focus trace out a limited parameter space in terms of PSF wing structure and speckle location.  Reference PSFs are {\em essential} for subtracting off the residual wings and speckles of the PSF from the coronagraph to obtain maximum contrast for JWST and HST.   Classical RDI further mitigates self-subtraction of point sources and axisymmetric low order structure in a sky scene that comes from post-processing techniques.  The highest contrast observations with smallest inner working angle in the optical from space achieve 10$^{-8}$ contrast/resel at $\sim$1\arcsec\ using a combination of ADI+RDI\cite{schneider14}.  Utilizing reference PSFs to enhance contrast may still be critical for CGI operations, especially if residual speckle patterns are at the level of 10$^{-8}$.  For example, images of the bright stars that are used to create the CGI dark hole should be retained to help construct a PSF library.

It is important to note that the ultimate metric in the context of first visits of an exo-planetary survey for WFIRST-AFTA CGI is not contrast, but False Positive Probability (FPP) (which if too large would trigger unnecessary follow-up observations and thus diminish the scientific return of the mission). In this context, obtaining two rolls, such as suggested for disk imaging by Ref. \citenum{Schneider-a-2014b}, is a key piece to quantifying FPP when using an approach similar to the one carried out with the HR8799 NICMOS data~\cite{Soummer-a-2011}. We thus recommend that this observing strategy be studied in the WFIRST-AFTA CGI, in spite of the possible operational complications associated with spacecraft rolls.  The reduction rate in FPP has not been explicitly studied and so its potential benefit is hard to quantify and is probably highly dependent on the specifics of the CGI design.  However, LOCI-processed NICMOS two-roll data of HR~8799 showed that even planets detected at SNR=3 in individual images can still be recovered versus similar brightness speckles due to two separate spacecraft orientations \cite{Soummer-a-2011}.

\subsection{Impact of the stability and the amount of instrumental aberrations on post-processing in the context of WFIRST-AFTA}\label{section-simus}

As we saw in the context of HST, the RDI strategy is confronted with the temporal stability of the instrumental aberrations, which leads to unsubtracted speckle residuals in the PSF subtraction process. 
A crucial issue is to understand to which extent the stability and the amount of aberrations will have an impact on post-processing in the context of WFIRST-AFTA. In this sub-section, we address the following questions: How will the stability of aberrations have an impact on post-processing in the context of WFIRST-AFTA? What will be the impact of the amount of aberrations? What could be the impact of the instrumental strategy on post-processing? 

\subsubsection{Simulations and reduction techniques}

To find answers to these questions, we considered some noise-free data sets simulated by the Jet Propulsion Laboratory (JPL) with their diffractive model for the WFIRST-AFTA coronagraph~\cite{Krist-p-2014} and from thermal models generated by the Goddard Space Flight Center (GSFC).  
We considered first and third iterations of thermal models (labeled as OS1 and OS3 by the WFIRST-AFTA project). The first one corresponds to a Fall 2014 model for which we artificially doubled and quadrupled the coma to study the robustness of post-processing. The second one is a Spring 2015 design for which we studied the performance with and without a LOWFSC subsystem. They all consist of the simulated consecutive observations of two stars: a bright star is targeted ($\beta$~Uma; A1IV, V=2.37) for the dark hole generation; it is then the science target's turn (47 Uma; G1V, V=5.04) to be observed. The coronagraph used to generate these data is the HLC.
There are 88 $\beta$~UMa images and 321 each for 47 UMa for the OS1 scenario and 8 $\beta$~UMa images and 17 each for 47 UMa for the OS3 scenario. Three synthetic planets were added to 47 UMa.
For a full discussion of these simulations, we invite the reader to refer to the paper of John Krist in the same JATIS volume.

We applied the classical PSF subtraction and KLIP techniques to reduce these data sets. 
Before reducing the data with KLIP, we subtract the average values of the science and reference data cubes so that they have zero mean. We collapse the science data cube (47 Uma) to obtain a long exposure raw image. 
We then apply the KLIP algorithm, which 1) decomposes the reference data cube into principal components or KLIP modes; 2) projects the long exposure raw image into these modes to create a synthetic reference; and 3) subtracts the synthetic reference from the target. 
The classical PSF subtraction consists of a simple subtraction of the collasped and normalized reference data cube from the normalized long exposure raw image. 

\subsubsection{Results and analysis}

\begin{figure}
  \centering
 \includegraphics[width = 13.cm]{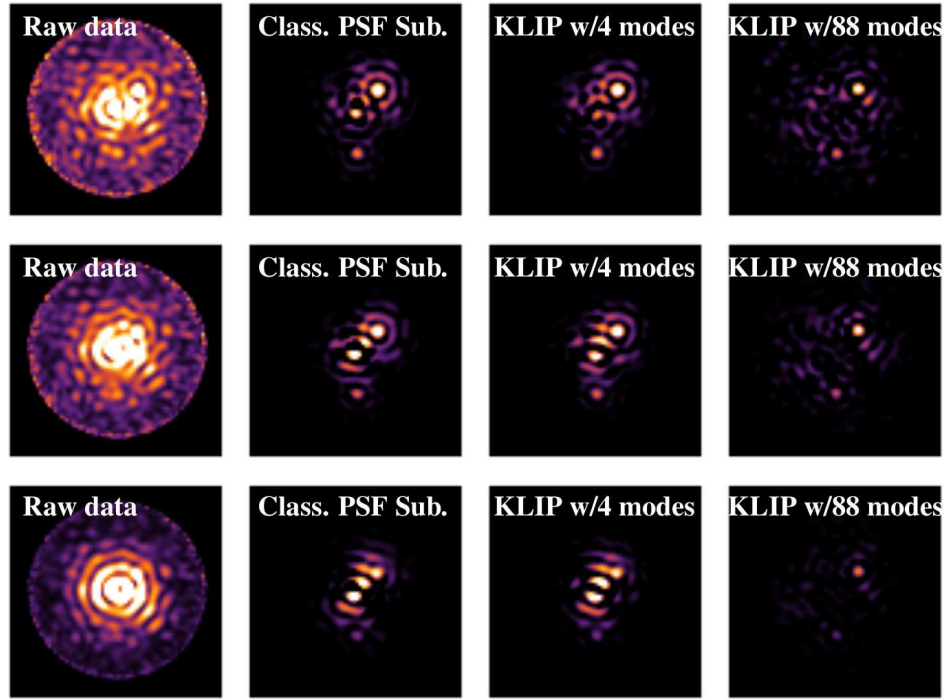}
    \caption{\label{Fig-os1_coma} HLC OS1 simulated data sets - Impact of the amount of aberrations on post-processing. Comparison of the noiseless raw data and three different reductions for increasing amounts of coma. Top: coma x1 (reference amount). Middle: coma x2. Bottom: coma x4. From left to right with same dynamic range: raw image (long exposure); image processed with classical PSF subtraction; image processed with KLIP 4 modes; and image processed with KLIP 88 modes. Whereas the two techniques show a relatively good and similar PSF subtraction compared to the raw image for the reference amount of coma (coma x1), the contrast improvement with 88 KLIP modes is better than with the classical PSF subtraction when the amount of coma increases.}
\end{figure}

\begin{figure}
  \centering
 \includegraphics[width = 13.2cm]{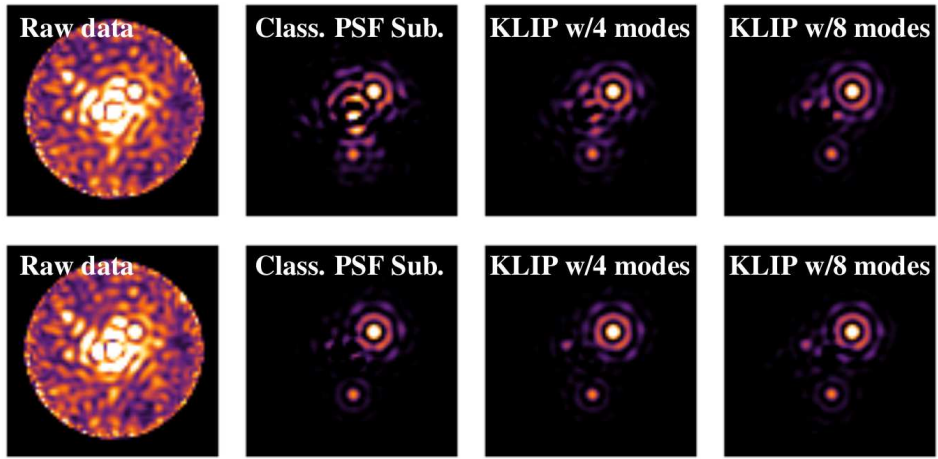}
  \caption{\label{Fig-Slide10} HLC OS3 simulated time series - Impact of the LOWFSC on post-processing. Comparison of the noiseless raw data and three different reductions. Top: data without LOWFSC. Bottom: data with LOWFSC. From left to right with same dynamic range: raw image (long exposure); image processed with classical PSF subtraction  image processed with KLIP 4 modes and image processed with KLIP 8 modes. The two techniques show a relatively good PSF subtraction compared to the raw images for both the cases without and with LOWFSC. The reduction quality is improved by the KLIP reduction compared to the classical PSF subtraction; the more KLIP modes used, the better the result.
}
\end{figure}

We compared the results of our reductions for the two simulated observing scenarios OS1 (Figure~\ref{Fig-os1_coma}) and OS3 (Figure~\ref{Fig-Slide10}) by showing each simulation with the same scaling: the long exposure raw image; 
the classical PSF subtraction reduction; the KLIP reduction with 4 modes; and with the maximal number of modes for OS1 (88 modes) and OS3 (8 modes). 
All the reductions show a relatively good PSF subtraction compared to the raw image, even with increasing amounts of coma or without LOWFSC. 

The performance of PSF subtraction appears to be sensitive to the thermal model of WFIRST/AFTA, especially for classical PSF subtraction.  The thermal model in OS1 leads to very stable wavefronts and under these stable conditions, the level of detection is quite similar in the three different reductions, with two of the three synthetic planets detected (Figure~\ref{Fig-os1_coma}.a). 
In the case of OS3, even though the number of data used to build the reference library is much smaller (8) than in the OS1 case (88), KLIP post-processing detects three planets instead of two with classical PSF subtraction (Figure~\ref{Fig-Slide10}.a).  This primarily is because the thermal model of OS3 leads to more unstable wavefronts than for OS1. In such conditions, the performance of classical PSF subtraction decreases and one should rather use KLIP as a reduction strategy.

Figure~\ref{Fig-os1_coma} gives insight into which extent the amount of aberrations has an impact on post-processing.
Even though the thermal model used for the simulations is the same in the three OS1 data sets, KLIP with 88 modes does a better job than the classical PSF subtraction at cleaning the residual speckles when the amount of coma increases. The case of coma multiplied by 4 (Figure~\ref{Fig-os1_coma}.c) is the most explicit as the KLIP reduction with 88 modes is the only one that enables the detection of two of the three synthetic planets.  
KLIP is more robust than the classical PSF subtraction to increasing amounts of aberrations, enabling very similar detections to the case with lower aberrations represented in  Figure~\ref{Fig-os1_coma}.a. 

Figure~\ref{Fig-Slide10} studies the impact of the LOWFSC, as instrumental strategy, on post-processing.
The LOWFSC provides enough aberration stability to improve the detection levels for the three different reductions in a similar way (Figure~\ref{Fig-Slide10}.b). 
Without LOWFSC, the detection level is better with KLIP (3 planets detected with 8 modes) than with classical PSF subtraction (2 planets detected) (Figure~\ref{Fig-Slide10}.a). 
Even the case with 4 KLIP modes enables a better detection, if not ideal, than the classical PSF subtraction for the third planet. 
An instrumental strategy such as the LOWFSC, aiming at stabilizing aberrations,  decreases the differences between the two reduction techniques, but KLIP could be of a great help if the LOWFSC does not work perfectly. 

As for HST, the stability of aberrations seems to have an important impact on post-processing in the context of WFIRST-AFTA. The performance of the PSF subtraction methods will highly depend on the thermal variations induced by the observing scenario, these variations being possibly mitigated by instrumental strategies such as the LOWFSC. 
If the temporal stability of WFIRST-AFTA PSFs will not be known until flight, it is important to confront each new observing scenario with various post-processing techniques. This will help to identify the best possible strategies at the very high contrast imaging levels enabled by WFIRST-AFTA.  

\section{James Webb Space Telescope Operations}
\label{sect:JWST}
The science operations of JWST coronagraphs will represent of significant step forward when compared to HST: the observing strategy will be designed from the start to accommodate modern data analysis methods that take advantage of PSF libraries, these algorithms will be included in the pipeline. Moreover the high sensitivity of these coronagraphs to low order aberrations requires accurate target acquisition routines whose functionalities are reminiscent of the WFIRST-AFTA CGI LOWFS.

JWST has a total of nine separate options for coronagraphy in the NIRCam \cite{green05} and MIRI instruments \cite{wright04}, which makes coronagraphic operations a more central part of the observaotry, much like for WFIRST-AFTA.  NIRCam possesses a focal plane that includes five separate masks, and can be combined with a variety of filters.  MIRI, on the other hand, features four focal plane masks paired with a single filter.  A difference between operations with JWST from HST, from the point of view of the user, will be a more significant amount of support with regards to designing observations during the different phases of the proposal process within the Astronomers Proposal Tool (APT) software and support with calibration and post-processing.  From the point of view of operations, more care is also needed in target acquisition and astrometry for JWST, especially given the tighter pointing requirements of the Four Quadrant Phase Masks (4QPM) on MIRI.  All of these operational aspects hold important lessons for WFIRST-AFTA, which will also be a mission with strong science drivers that include coronagraphy as well as a significant high contrast imaging GO component.  In this Section we will review operational aspects of JWST that may have a bearing for CGI operations, including obtaining un-occulted photometric images, using dithering strategies to enhance contrast for some coronagraphic modes, and the development of data products and user tools that enable use of coronagraphic modes from a broader subset of the astronomical community. 

\subsection{Wavefront Sensing and Control}

As a deployable segmented telescope, JWST relies crucially on active control to achieve and maintain the required wavefront quality, using sensing methods based around focus-diversity phase retrieval using image data from the science instruments \cite{acton12}. While the desired control precision for JWST is less demanding than for WFIRST CGI, there are still lessons which are transferrable. In the broadest terms these reflect the fundamental point that developing the required control algorithms is only the first step in achieving a working practical system. Wavefront control algorithms should not be developed in isolation, but rather should from the beginning bear in mind the needs of efficient, cost-effective optical integration and test, and timely post-launch commissioning and calibration.  These mission phases are equally critical as the routine orbital science operations that are the primary focus of many design studies. 

A significant early design decision for JWST was that all wavefront analyses happen on the ground at the mission operations center, with control decisions made interactively by a human operator, and commands subsequently uplinked to the spacecraft on the next communications pass. This adds substantial operational complexity, imposes a minimum turnaround time, and thus limits the achievable temporal control bandwidths. Given the expected stability levels for JWST, this suffices.  For a notional WFIRST in geostationary orbit with continuous communications such an architecture might be acceptable, but for a WFIRST at L2 with periodic communications, it seems strongly desirable to place as much control authority as possible in the spacecraft itself. This would maximize the control bandwidth and allow generation of dark holes in times when the spacecraft is not in communication with Earth.  Enabling this may set significant requirements on spacecraft computing power and software complexity, requiring for instance an image calibration pipeline for the CGI imager to be resident on the spacecraft as well as the phase retrieval calculations themselves.

For observatory commissioning, the high fidelity and rapid wavefront sensing enabled by the planned CGI LOWFS could provide an efficient tool for sensing the aberrations of the active 2.4 m telescope and aligning the movable secondary post-launch. In the case of JWST, commissioning the active optics and aligning the segmented primary is expected to take $\sim$3 months, half of the overall observatory commissioning period. The JWST telescope commissioning plan today is significantly more complex than originally envisioned: over time the process has accreted additional steps to accommodate interactions with the spacecraft attitude control system, boresight pointing calibrations, measurement of influence functions for observatory thermal distortion, measurement and tuning of reaction wheel and cryocooler induced pointing jitter, and more. Further complexity was added due to needs for looping back to repeat earlier steps after correcting for certain mirror influence function terms and removing degeneracies.  Most of these practical details fall outside a narrow definition of ``wavefront control'', but all have proven necessary in practice to achieve the desired RMS wavefront error for JWST.  While WFIRST does not have a segmented primary the required control levels are much tighter, and it will require active alignment of its optical telescope assembly to achieve Hubble-like contrasts prior to engaging the deformable mirrors. Planning for the commissioning process to go from a post-launch telescope in an unknown state to a dark hole at 10$^{-9}$ or better contrast should be considered in detail as soon as possible, including realistic estimates for practical complications and an accounting for observatory overheads. 

Closely related is the challenge of testing and validating the wavefront control system pre-launch. JWST has relied on a combination of a laboratory testbed \cite{acton06} and integrated software modeling \cite{knight12} to validate the control processes. Even with tremendous expense and effort using the world's largest cryo-vacuum chamber, limitations of the test environment on the ground due to vibration and gravity sag prevent truly achieving ``test as you fly''. The ability to conduct even limited end-to-end tests of wavefront control in closed loop with the flight hardware occurs relatively late in the integration and test program. Testing and development of the flight software relies almost entirely upon simulations and mock data, but the simulator software cannot itself be fully validated until equally late in the integration and test stage.  The complexity is compounded by the fact that relevant software and hardware components of the wavefront sensing \& control system have been developed by several different project partners, and the wavefront sensing subsystem relies on and interacts with every other subsystem of the ground system; this necessitates much further time-consuming systems integration, and again depends heavily on simulations rather than real instrument test data.  Close collaboration between various project teams has been necessary to successfully integrate these complex subsystems and develop a comprehensive test program within the constraints of the ground environment. For WFIRST, careful attention to systems architecture and integration is needed to develop a practical and cost-effective development and test plan. Is it possible in a meaningful way to demonstrate the complete system on the ground pre-launch using the flight hardware to achieve ultra-high contrast in a suitable test chamber?  If not then WFIRST must, like JWST, rely heavily on complex modeling to validate performance and control processes.  In that case, an overall roadmap for wavefront sensing test and commissioning should be developed as soon as possible. Such a plan should consider in a unified way both ground test and post-launch commissioning of WFIRST, and consider what steps and cross-checks will best mitigate risks and provide the necessary learning experience with the flight hardware to enable rapid commencement of science operations post-launch.

\subsection{Un-occulted Images for Photometry}
Successful coronagraphic observations remove the light from a host star from the image, but there are times when information about the host star is desirable.
For example, direct photometry of target stars with JWST in its coronagraphic modes will be important both from scientific and operational points of view.  Scientifically, the photometry of the central star is useful for non-detection analyses.  Non-detections still constrain the orbital architecture of planetary systems in which one planet is already discovered, or in the case of large populations of stars that have been blindly searched for planets.  Photometry can also determine the intrinsic variability of a source, which is particularly important for young stars that are often variable at non-negligible levels.  Addtionally, contrast limits provide utility for the planning of future observations based on true CGI performance.  On the operational side, photometry will allow for direct scaling to reference PSF stars within a given bandpass and will allow for the monitoring of instrument performance throughout the mission lifetime.

\subsection{Enhancing Contrast and Efficiency}
To ensure the best contrast performance in space with HST and JWST, a reference PSF star must be observed close in time (or telescope thermal/optical state) to the science target and spacecraft orientation changes must be performed for the target as well (i.e, as described in Section \ref{sec:psfsub}.  Additionally, for coronagraphic modes where pointing accuracy is below the requirement for best contrast, such as the 4QPM on MIRI\cite{cavarroc08}, additional operational techniques need to be employed.  

The target acquisition process of JWST described in Section \ref{sec:JWSTastro} does not mitigate the fact that the repeatability of small angle maneuvers and target acquisition for JWST has a non-negligible impact on contrast for high contrast imaging modes that are particularly sensitive to misalignments \cite{lajoie14}.  Most coronagraph designs are sensitive to misalignments between the mask and target.  Using RDI subtraction also requires a similar placement of the reference behind the coronagraphic mask relative to the initial science target placement.  One solution to this problem is to employ sub-pixel dithering scheme to sample the diversity of PSF structure under the action of sub-pixel mis-alignments.  These sub-pixel dithers, in concert with post-processing techniques such as LOCI or KLIP, can enhance delivered contrast for JWST coronagraphs by factors of a few to 10\cite{soummer14}.  Sub-pixel dithering can also be used for the CGI, depending on how large non-repeatability is within the instrument relative to its ultimate contrast sensitivity to mis-alignments.  Similar methods are currently being investigated for the BAR5 location on STIS as well, where contrast is limited by how well the target star and its PSF reference are centered behind the occulter.

\subsection{Data Products and User Tools}
\label{sec:dput}
Beyond the main mission of the WFIRST-AFTA CGI, an infrastructure will need to be in place to maximize the science return from GO programs. In order to do that, general users at varying levels of expertise should be able to seamlessly plan observations and work with data once it's been executed.  Operationally, this means a robust infrastructure for calculating exposure times for various coronagraphic modes, tools to design observing sequences, clear policies for use of the CGI, a pipeline that generates easily analyzed products, and additional resources to maximize the science impact of a given set of observations at the end of the mission.  These issues have undergone extensive planning for JWST as well. 

For most coronagraphic observations on JWST, the largest term in the noise budget is associated with systematic errors, such as residual speckles, that do not average out when the integration time increases. As a consequence, speckles need to be subtracted from the science exposure in order to improve the SNR of the sources observed (often very faint when compared to the parent source under the coronagraphic mask). This means that observers need to obtain calibration images to subtract systematics from the science images. This in turns drives a series of specific use-cases, which have to be taken into account when designing the observation planning and support tools, such as an exposure time calculator (ETC).  We summarize the current recommendations for JWST's coronagraphic ETC\cite{jwst003862}, representing the first mission to invest in specific tools for coronagraphic planning.  The JWST ETC will need to support the same coronagraphic observing sequences and strategies that will be defined in the observation planning phases.  Primarily this is through accurate estimates on the photon noise associated with the PSF wings from JWST and the associated systematic speckle noise that occurs after RDI subtraction.  Fairly sophisticated modeling of the 2-D JWST PSF is thus required for various modes, as well as a good model for the speckle noise floor associated with each mode and observing strategy.  Finally, the observatory performance will need to be monitored to update these values as the mission progresses.  A similar strategy will make sense for the CGI as well, where 2-D speckle noise characteristics as a function of mode will be the driver for signal-to-noise calculations of potential targets. 

Coronagraphic observations are typically very complex, which means that efficient observation planning is desirable.  The strategy that will likely be employed for JWST in concert with the Astronomer's Proposal Tool (APT) uses the concept of both instrument-level templates and of a higher-level ``super-template'' for coronagraphic observations\cite{jwst004141}. An instrument-level template will be a simple coronagraphic sequence for a single target.  

In the case of JWST, a basic template will include observing the target, executing a spacecraft roll, observing the target a second time, and then observing a reference star, along with other options such as for sub-pixel grid dithering\cite{soummer14}.  This sequence is born primarily from studies on the most efficient observing set-up for the broadest set of use-cases for coronagraphy (i.e., Ref. \citenum{jwst004140} for use cases and Ref. \citenum{jwst004165} for an efficiency study).  A similar approach should also be followed for the CGI, especially with regards to designing an efficient Design Reference Mission (DRM) while minimizing overheads.  Efficiency studies are particularly salient given real scheduling constraints in the proposed geosynchronous orbit of WFIRST-AFTA versus long exposure times for exoplanet characterization.  

At a higher level than a basic template, JWST will also utilize super-templates to collect multiple observing sequences into surveys or collections of targets. This approach accommodates the basic needs for the majority of users and science cases, while preserving sufficient flexibility for customization to accommodate particular science cases, especially for expert users. Current unknowns about the JWST observatory performance also motivate this approach, including slew accuracy and thermal temporal evolution, which impact target acquisition, wavefront error, and coronagraphic PSF.

JWST's operational approach for coronagraphic science products is to implement a pipeline that delivers nearly final products, implementing classical RDI subtraction in conjunction with co-adding individual exposures that were taken at separate dither positions behind the coronagraphic mask or through spacecraft orientation changes.  Additionally, the goal is to also include some level of post-processing while utilizing a public reference PSF library similar in style to the ALICE project that uses KLIP \cite{soummer12,choquet14}.  Final products not only will include reduced images, but several other metrics common to high contrast observations, including S/N maps and contrast performance metrics.  This represents a shift in philosophy relative to current aspects of high level coronagraphic science products with HST that have generally not been directly supported in the standard data pipelines of ACS, STIS, or NICMOS but that have primarily been executed through the funding of legacy archival proposals such as Legacy Archive PSF Library And Circumstellar Environments (LAPLACE HST-AR- 11279, PI: Schneider) and ALICE.  This has led to a generally heterogeneous coronagraphic dataset where each PI may have chosen slightly different approaches to high contrast imaging.  JWST's operational approach will be more standardized, but with a nod to retaining flexibility for experimentation.  One key component of this type of pipeline is that the policies for coronagraphic observing programs will require that reference PSFs be released publicly with no proprietary period, and justification will need to be given if an observing plan deviates from the standardized coronagraphic approach.
 
Leveraging existing observational planning infrastructures, such as through JWST's ETC or through the tools related to other past missions, will be essential for maintaining the rapid pace of development for WFIRST-AFTA.  Since the WFIRST-AFTA CGI has a large fraction of its time devoted to its prime mission, it will also be useful to follow a standardized approach to better construct a homogenous final high contrast dataset, while still generating a large library of reference PSFs that could potentially be used both for post-processing of the main survey and for use in GO programs.  An opportunity for constructing such a library will exist through stars with no dust or companions in the main survey as well as short exposures on the bright stars that are used for wavefront control activities.  Short exposures with no dark hole may also be desirable to allow post-processing for programs that do not require a high contrast dark hole, such as nearby debris disk observations whose angular size is larger than 1-2\arcsec.
 
 \section{Target Acquisitions and Astrometric Precision}
 \label{sec:JWSTastro}
 Successful exoplanet observation and characterization with the CGI will require accurate target acquisitions and stable pointing.  The expectation is that stars will be placed on the CGI detector with an initial accuracy of $\sim$1\arcsec, placed behind the coronagraphic spot via the FSM, and the overall spacecraft jitter (and any drifts to the instrument) will be corrected via the LOWFS.  
 In reality, there is always some raw uncertainty in the position of the target star on the sky, convolved with the inherent uncertainty of where the spacecraft's Fine Guidance System has placed the target.  The optimal center behind the mask must also be determined during commissioning.  For both JWST and HST, target acquisition procedures are handled by each instrument, especially where positioning on the instrument's detector is important for a given observing mode.  The final accuracy is obtained by a combination of initial target coordinate accuracy requirements, as well as requirements on the accuracy of a given target acquisition algorithm used within the flight software.  Both the HLC and SPC are sensitive to jitter, and the HLC is particularly sensitive to where the target is placed behind the mask.  Therefore, much like HST and JWST, there will need to be a way to hand off the target from the initial FGS pointing to the optimal location behind a mask, at least for the HLC. 

HST target acquisition procedures primarily use centroiding within a subarray of an instrument's detector to locate the source and then execute a small angle maneuver to place the target in the appropriate location.  STIS' target acquisition for its coronagraphic aperture is identical for all aperture positions--a coarse location first occurs in a 5\arcsec$\times$5\arcsec\ subarray where the science target's centroid is measured and a small angle maneuver is calculated based on a look-up table of supported aperture locations \cite{downes97}.  A second image of the object is obtained, and then the slit wheel is rotated into place.  A lamp illuminates the focal plane and the flight software executes a calculation to determine the exact position of the aperture on the detector.  A final small angle maneuver is executed to place the target behind the aperture.  Additional steps are currently in place for the bent finger occulter on STIS.  There, two dithers of $\sim$12~mas are performed for both the target and the reference PSF star in an effort to mitigate the 13~mas placement uncertainty between them and maximize contrast performance.  The uncertainty is limited by a combination of the target acquisition and the slit wheel non-repeatability.

The procedure for NICMOS was relatively similar to the basic target acquisition of STIS, but additional imaging of the corongraphic hole was implemented to derive the best placement of a target \cite{schultz98,schultz99}.  The result was an acquisition accuracy of 5.9~mas to the coronagraphic hole center, with repeatability orbit-to-orbit of 2.9~mas \cite{schultz04}.  

 JWST may be utilizing un-occulted images in the case of the NIRCam coronagraphs to define a local astrometric reference and aid in determining the target star location behind a coronagraphic mask with a goal of 5~mas precision in astrometry\cite{jwst004166}.  For NIRCam, target acquisition images are first obtained through a neutral density filtered subarray in an intermediate focal plane before the detector; the target is then placed behind a mask with a small angle maneuver.  Full frame images are then obtained both during the acquisition and after the target is occulted to create the astrometric reference and directly measure the magnitude of the small angle maneuver from the image, rather than from spacecraft telemetry alone.  A similar approach could be applied to the WFIRST-AFTA CGI: if the offset between the wide field imager (WFI) of WFIRST-AFTA and the CGI is precisely known (to better than 5-10~mas), parallel WFI images could serve to lock in the astrometric reference and anchor the target star position within the CGI.  
 
 Most of these approaches can be applied to the CGI in concert with fiducial copies of the occulted source outside of the dark hole region to provide photometric and astrometric information, which is part of the current CGI deformable mirror operational design \cite{sivramakrishnan,marois}.  Assuming that the optimal location for a star behind the mask is defined and monitored with time, the centroid of the occulted star can be measured and then placed behind the mask.

 \begin{figure}
   \begin{center}
   \begin{tabular}{c}
   \includegraphics[width=10cm]{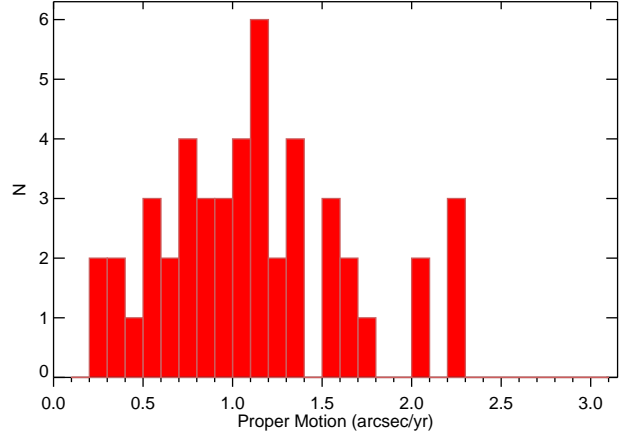}
   \end{tabular}
   \end{center}
   \caption 
   { \label{fig:f4} Distribution of proper motions for the 50 closest stars in the HabCat as defined by Reference \citenum{turnbull03}. The star with the smallest proper motion in this sample is HIP~23452, which has a total proper motion vector of 263 mas/yr} 
   \end{figure}

In addition to an accurate knowledge of the central star, astrometric precision requirements for the CGI are driven by the need to quickly follow-up and confirm a candidate's common proper motion with the central star at the 3-5$\sigma$ level.  The astrometric requirement (under the assumption that the precision is the same for measurements along one direction of the detector and the orthogonal direction) is:

\begin{equation}
\sigma_{\rm ast}=\frac{|\mu_{\star, {\rm min}}|\Delta t}{7}
\end{equation}

where $\sigma_{\mathrm ast}$ is the astrometric precision requirement, $\mu_{\star,{\rm min}}$ is the minimum proper motion observed, $\Delta t$ is the time between observations, and the factor of seven comes from the requirement of a 5-$\sigma$ detection of common proper motion and a factor of $\sqrt{2}$.  Several previous discoveries (and refutations) of common proper motion candidates required accuracy at the level of 5-10~mas \cite{rameau13}.  Figure \ref{fig:f4} presents a possible scenario for the WFIRST-AFTA CGI based on a blind search of the 50 closest stars in the HabCat \cite{turnbull03}, which is a rough proxy for a blind search target sample that is somewhat biased towards nearby FGK stars that do not have resolved binary companions.  We assume a three month interval between observations to determine the presence of common proper motion for a candidate.  The magnitude of the smallest proper motion of this sample is 263~mas/yr, so the minimum $\sigma_{\rm ast}$7~mas for a minimum follow-up time of 3 months.

Measuring significant orbital motion of the companion is also necessary, especially for follow-up observations and mass determinations of radial velocity discovered planets.  Predicting the precise orbital motion of an exoplanet is beyond the scope of this paper, so we consider two idealized cases that should roughly give us a range of orbital motion--the face-on circular orbit of an exoplanet in orbit around a sun-like star at the IWA of the CGI at 10~pc (Case I; a=1~AU, P=1~yr) and a similar orbit at the edge of the dark hole at 30~pc (Case II; a=55~AU, P=410~yr).  Using the above three month interval between observations, the magnitude of the difference in position for the companion will be 141~mas for Case I and 7~mas for Case II.  Case I will complete multiple orbits over the WFIRST-AFTA nominal mission lifetime of 5~yr, while Case II will travel roughly 140~mas at the end of the mission.  Astrometric precision on the order of 5-7~mas would provide $>$5-$\sigma$ orbital motion detections for the majority of planets within the first year of detection.  A similar accuracy is required for precisely determining the orbit of an exoplanet with a known radial velocity semi-amplitude, which will allow for accurate exoplanet mass determinations \cite{brown15}.

Astrometric precisions of 5~mas would also be desirable to broaden the application of exoplanet detection/characterization with the CGI to the nearest star forming regions at 100~pc.  Optimization of the actual exoplanet search strategies for WFIRST-AFTA is ongoing as part of work determining a design reference mission \cite{traub,spergel15}, but multiple visits for blind searches might be required not only to confirm proper motion, but to disentangle companions from background objects\footnote{http://exep.jpl.nasa.gov/stdt/Exo-S\_Starshade\_Probe\_Class\_Final\_Report\_150312\_URS250118.pdf}, surrounding dust structures \cite{starkpseudozodi,defrere12}, or to maximize planet yields due to changing planetary phase curves \cite{starkayo}.

\section{WFIRST-AFTA Operations to support polarimetry}
The polarization of light from astrophysical objects is a powerful diagnostic tool that can provide information about the nature of the object that cannot be gleaned otherwise. For circumstellar environments, linearly polarized light results from starlight that is scattered from dust particles surrounding the star and from the atmospheres or surfaces of planets and small bodies. In combination with a coronagraph, polarimetry can thus yield information about the circumstellar material not provided by coronagraphic imaging on total light alone.  For example, observations of circumstellar debris disks have shown fairly high polarizations of a few tens of percent, and yield important insight into the scattering material, the physical structure and the geometry of the system (e.g., Refs.\citenum{perrin09,thalmann13,perrin14}).  In addition to the diagnostic power of polarimetry, the fact that these debris disks are highly linearly polarized can be exploited to enhance the contrast achieved (by factors of $\sim 10x$), because the central starlight is intrinsically unpolarized, so will null when the polarized intensity is measured (see, e.g., \citenum{perrin15}) . Given these significant advantages to improving the science return from a corongraphic system, we next explore the possibility of including such a system in the CGI design, and its implications for operations. We also present some important lessons learned from previous space-based deployment of polarimetry enabled coronagraphs on {\it HST}.

In order to measure the full linear Stokes parameters ({\it I, Q, U}), and thus the degree of linear polarization (p) and the position angle of the electric vector on the sky ($\theta$), the polarized signal must be measured at least three times with the beam analyzer rotated by 60$^\circ$ or 120$^\circ$; this rotation can also be accomplished by placing a rotating half-waveplate in front of the analyzer, or by rotating the analyzer itself, which for a fixed system translates into rotation of the instrument or telescope.

The current baseline CGI does not include a formal polarimetry mode. However, the design does include a single linear polarization analyzer, which is used in the coronagraphic imaging channel to mitigate loss of contrast caused by intra-pupil polarization mixing as a result of the fast optical system\cite{spergel15}.  By selecting only one of the two orthogonal polarization states of the incoming beam, these effects are minimized. However, this technique does introduce other problems for measuring the total intensity of objects, especially if their emission is polarized\cite{schneider15}.

Operationally, with the currently defined CGI design, this would require observing an object at three different spacecraft roll angles.  Observations at multiple roll angles have already been discussed in terms of the ADI technique, so this would not entail a major change in overall operations, except for the requirement that the roll angle be 60$^\circ$ or 120$^\circ$. Since the {\it WFIRST-AFTA} observatory is restricted to rolls of $\pm 15 ^\circ$,  such large rolls are not possible at a single ``visit'' to the target. However, as for {\it HST}, they can be accomplished by observing the target at different times of the year. This requirement may cause restrictions on scheduling of observations and could eliminate some targets.  Even though not currently base-lined, slight design changes such as the addition of a half-waveplate to the CGI as explored in an accompanying article in this special issue \citenum{schneider15}, could enable a polarimetry mode without the need for spacecraft rolls. These changes would still preserve the mitigating effects of the single analyzer design, while opening up important new science capabilities. 

Space-based coronagraphic polarimetry has been been implemented with the {\it HST}/NICMOS and  {\it HST}/ACS instruments \cite{hines00, hines07, graham07, maness09,perrin09,perrin14}.  Since most of our experience with such systems is with NICMOS, we will restrict our discussion to it, noting that similar operational considerations apply to the ACS.

The NICMOS filter wheel contains three POLARCOR polarizing filters sandwiched with 2.0 $\mu$m band-pass filters\footnote{Three other polarizers are sandwiched with $\approx 1.1\mu$m pass-band filters, but these are only available for the NIC1 camera, which does not have a coronagraph.}, and offset in position angle by $\approx 120^\circ$. Full linear Stokes parameters can then be obtained by imaging the object in all three polarizers, or by obtaining images through a single polarizer with the telescope rolled to three different angles differing by 60$^\circ$ or 120$^\circ$.

Ideally, the polarizers would all have the same polarizing efficiencies and have position angle offsets differing exactly by 120$^\circ$, but unfortunately this was not realized in the final as built instrument\cite{hines00}.  Operationally, this required careful calibration of the instrument on the ground and on-orbit to determine the proper coefficients to transform the three images into full Stokes images \cite{hines00,batcheldor06,batcheldor09}, and high precision polarization measurements ($\sigma_p \le 1\%$) are routinely obtained. We note that the lessons learned from NICMOS were heeded by the ACS designers, resulting in an ACS polarimetry mode that was much closer to ideal than NICMOS, and thus required less intensive calibration to achieve similar polarization precisions.

For  {\it WFIRST-AFTA}, the baseline CGI design will require observations of polarized and unpolarized sources to characterize the polarimetry optics, so additional calibration of an enhanced polarimetry mode would not be a major perturbation on the system. In addition, the possible design enhancement suggested by \citenum{schneider15} would use only a single analyzer and a half-waveplate, so problems with polarizing efficiency difference between polarizers are not present. In addition, because the rotation angle of half-waveplate can be adjusted to high precision with simple, flight-proven mechanisms, non-ideal offsets  in the position angle can be eliminated.  

\section{Conclusions}
The WFIRST-AFTA/CGI will be a large leap forward in the quest for directly imaging nearby terrestrial planets.  The technology development for this instrument will create a strong infrastructure that supports the long-term goals of NASA, namely a planet surveyor that not only detects terrestrial planets within a few AU of nearby stars but also characterizes their atmospheres\cite{spergel15}.  Operational concepts must also evolve in concert with great technological leaps so that the {\em maximal} science return is realized for a given instrument for the broadest cross-section of the user community.  In this paper we have identified some operations that may translate well with the WFIRST-AFTA/CGI, and additional aspects that may impact the expected performance of such an instrument.

We find that optical detectors on the CGI will need to be monitored carefully for degradation due to radiation damage in particular, which could impact the ultimate sensitivity to faint sources over the full mission lifetime.  A high level of repeatability (below 5~mas) will mitigate many risks to the CGI's performance, both in terms of the mask wheels in the current CGI design as well as in terms of the absolute target acquisition and pointing accuracy.  Some of these error terms will be mitigated by operation of the LOWFSC, but additional calibration images will be desirable to construct local astrometric frames on the CGI detector or in concert with the WFI.  One of the other operational tall poles will be to design a coherent and effective observing template for CGI that maximizes raw contrast on the telescope, but with an eye toward boosting the success of various post-processing techniques executed through calibration pipelines.  Wrapped around the telescope operations are the complementary activities for enabling clear observation planning and data analysis for a general observer community that will make up 25\% of observing time on WFIRST-AFTA, despite its many focused science goals.  A broad appeal to a wide swath of the observer community will also push the boundaries of the CGI performance and hopefully spur further innovation in high contrast imaging techniques.

%%%%%%%%%%%%%%%%%%%%%%%%%%%%%%%%%%%%%%%%%%%%%%%%%%%%%%%%%%%%%
\acknowledgments 
The authors would like to acknowledge the careful and painstaking work of A. Gaspar and G. Schneider in their commissioning of the bent finger occulter on STIS, the results of which were re-analyzed and presented in Figure \ref{fig:f1}.  J. Taylor was instrumental in maintaining the monitoring code for STIS dark rates that were used to create Figure \ref{fig:f2}.  Support for this work was provided
by the WFIRST Study Office at the NASA Goddard Space Flight Center
(GSFC), as part of joint preformulation science center studies by the
Space Telescope Science Institute (STScI) and the Infrared Processing
and Analysis Center (IPAC).  The work reported in Section \ref{section-simus} was carried out in part under sub-contract 1506553 with the Jet Propulsion Laboratory funded by NASA and administered by the California Institute of Technology.
 B. Mennesson and J. Krist for providing simulated data and helpful discussions used in that section.

%%%%%%%%%%%%%%%%%%%%%%%%%%%%%%%%%%%%%%%%%%%%%%%%%%%%%%%%%%%%%
%%%%% References %%%%%

\bibliography{report,mygouf,report_mygouf}   %>>>> bibliography data in report.bib
\bibliographystyle{spiejour}   %>>>> makes bibtex use spiejour.bst

%%%%%%%%%%%%%%%%%%%%%%%%%%%%%%%%%%%%%%%%%%%%%%%%%%%%%%%%%%%%%
%%%%% Biographies of authors %%%%%

\vspace{2ex}\noindent{\bf John Debes} is an ESA/AURA astronomer at the Space Telescope Science Institute.  He received a BA in Physics in 1999 at the Johns Hopkins University and a PhD in Astronomy and Astrophysics from the Pennsylvania State University in 2005.  He has authored or co-authored 40 published papers and one book chapter.  His interests include high contrast imaging of exoplanets and debris disks, as well as the study of planetary systems that survive post-main sequence evolution.

\vspace{2ex}\noindent{\bf Marie Ygouf} is a Postdoctoral Researcher at STScI. She received her B.S. in Fundamental Physics from the Universit\'e{} Pierre et Marie Curie in 2006, a M.S. degrees in Optics and Photonics from the Institut d'Optique Graduate School in 2009 and her Ph.D. in Astrophysics from the Universit\'e{} de Grenoble in 2012. Her research focuses on improving the performance of high contrast imaging instruments, particularly by developing innovative image post-processing techniques.

\vspace{2ex}\noindent{\bf Elodie Choquet} is a postdoctoral fellow working on postprocessing techniques for high-contrast imaging. She is part of the ALICE project, which aims at re-analysing the NICMOS coronagraphic archive. She is also interested in wavefront control techniques and contributed to the design of the HICAT experiment which aims at studying wavefront control strategies for high-contrast imaging instruments on future complex-aperture telescopes. She defended her PhD in 2012 working on the fringe tracking algorithms for VLTI-GRAVITY instrument.

\vspace{2ex}\noindent{\bf Dean C. Hines} is a Scientist at the Space Telescope Science Institute (STScI). He uses visible and infrared (high-contrast) imaging, spectro- and imaging polarimetry, spectroscopy, and radio imaging to investigate active galaxies, quasars, stellar evolution, and the formation and evolution of planetary systems. He is the JWST/MIRI Team Lead at STScI, and Deputy PI for the proposed Exoplanetary Circumstellar Environments and Disk Explorer (EXCEDE). 

\vspace{2ex}\noindent{\bf Johan Mazoyer} is currently a postdoc at the Space Telescope Science Institute. He graduated from the \'E{}cole polytechnique (Paris, France) in 2011 and received a PhD in Astronomy and Astrophysics \textit{cum laudae} from Paris Diderot University/Paris Observatory (France) in 2014. He has authored or co-authored more than 15 publications. His research interests lie both in the development of innovative instruments for imaging close circumstellar environments (planets or dust) and in the analysis of high contrast images.

\vspace{2ex}\noindent{\bf Marshall D. Perrin} is an associate astronomer at the Space Telescope Science Institute, and a member of the telescope optics team there. His research interests focus on the development of advanced instrumentation and data processing methods for the characterization of extrasolar planetary systems, including adaptive optics, coronagraphy, integral field spectroscopy, and differential polarimetry. He received his PhD from the University of California, Berkeley, and was a postdoc in the UCLA Infrared Lab before joining STScI.

\vspace{2ex}\noindent{\bf Roeland van der Marel} is an astronomer at the Space Telescope Science Institute (STScI) and an adjunct professor at nearby Johns Hopkins University. At STScI he is the Mission Lead for WFIRST-AFTA. He is a frequent user of the Hubble Space Telescope, and an expert on black holes and the structure of galaxies.

\vspace{1ex}
\noindent Biographies and photographs of the other authors are not available.

\listoffigures
%\listoftables

\end{spacing}
\end{document}